\documentclass[twocolumn,prl,aps,showpacs,superscriptaddress]{revtex4-1}

\usepackage{siunitx}
\usepackage{graphicx}
\usepackage{amsmath}
\usepackage{float}
\usepackage{color}
\usepackage{mathrsfs}

\begin{document}
%%Title of paper
\title{Temperature-induced increase of spin spiral periods}
\author{Aurore Finco}
\email{aurore.finco@physnet.uni-hamburg.de}
\affiliation{Department of Physics, University of Hamburg, D-20355 Hamburg,
  Germany}
\author{Levente R\'{o}zsa}
\affiliation{Department of Physics, University of Hamburg, D-20355 Hamburg,
Germany}
\affiliation{Institute for Solid State Physics and Optics,
  Wigner Research Centre for Physics, Hungarian Academy of Sciences,
P.O. Box 49, H-1525 Budapest, Hungary}
\author{Pin-Jui Hsu}
\author{Andr\'{e} Kubetzka}
\author{Elena Vedmedenko}
\author{Kirsten von Bergmann}
\author{Roland Wiesendanger}
\affiliation{Department of Physics, University of Hamburg, D-20355 Hamburg,
Germany}
\date{\today}
\begin{abstract}
  Spin-polarized scanning tunneling microscopy investigations reveal
  a significant increase of the magnetic period of spin spirals in
  three-atomic-layer-thick Fe films on Ir(111), from about \SI{4}{\nano\meter}
  at \SI{8}{\kelvin} to about \SI{65}{\nano\meter} at room temperature.
  We attribute this considerable influence of temperature on the magnetic length
  scale of noncollinear spin states to different exchange interaction
  coefficients in the different Fe layers. We thus propose a classical
  spin model which reproduces the experimental observations and in which the
  crucial feature is the presence of
  magnetically coupled atomic layers with different interaction strengths. This
  model might also apply for many other systems, especially magnetic
  multilayers.
\end{abstract}
\pacs{}
\maketitle
%
% Introduction
Recently, significant attention is turned towards the possible spintronics
applications of noncollinear magnetic configurations in ultrathin films,
including chiral domain walls~\cite{Parkin} and isolated skyrmions~\cite{Fert}.
The Dzyaloshinsky--Moriya interaction~\cite{Dzyaloshinsky, Moriya},
induced by the breaking of inversion symmetry at the interface and the strong
spin--orbit coupling in the substrate, plays an important role in stabilizing
such noncollinear states. Crucial points for improving spintronics devices
do not only include decreasing the electrical currents and magnetic fields
necessary for their operation, but also ensuring a good thermal stability.
A possible solution to the latter problem is to use magnetic films or
multilayers with a thickness such that the critical temperature is well above
\SI{300}{\kelvin}, which has been proven successful in stabilizing
room-temperature skyrmions~\cite{moreau-luchaire_additive_2016, chen_room_2015,
  jiang_blowing_2015, boulle_room-temperature_2016}.
Here we aim to improve thermal stability in simple epitaxial systems by
increasing the thickness of an Fe layer on Ir(111), where the Fe/Ir interface
is known to induce a strong Dzyaloshinsky--Moriya
interaction~\cite{heinze_spontaneous_2011}.
Spin-polarized scanning tunneling microscopy (SP-STM) measurements
revealed a nanoskyrmion lattice in the monolayer Fe on Ir(111) which vanishes at
\SI{28}{\kelvin}~\cite{sonntag_thermal_2014}. Since it was
predicted and observed for various transition metal ultrathin films that the
Curie temperature increases with the thickness of the
film~\cite{jensen_calculation_1992, schneider_curie_1990, elmers_magnetic_1995},
we expect an improved thermal stability for the double- and triple-layer
Fe films.

\begin{figure*}
  \includegraphics{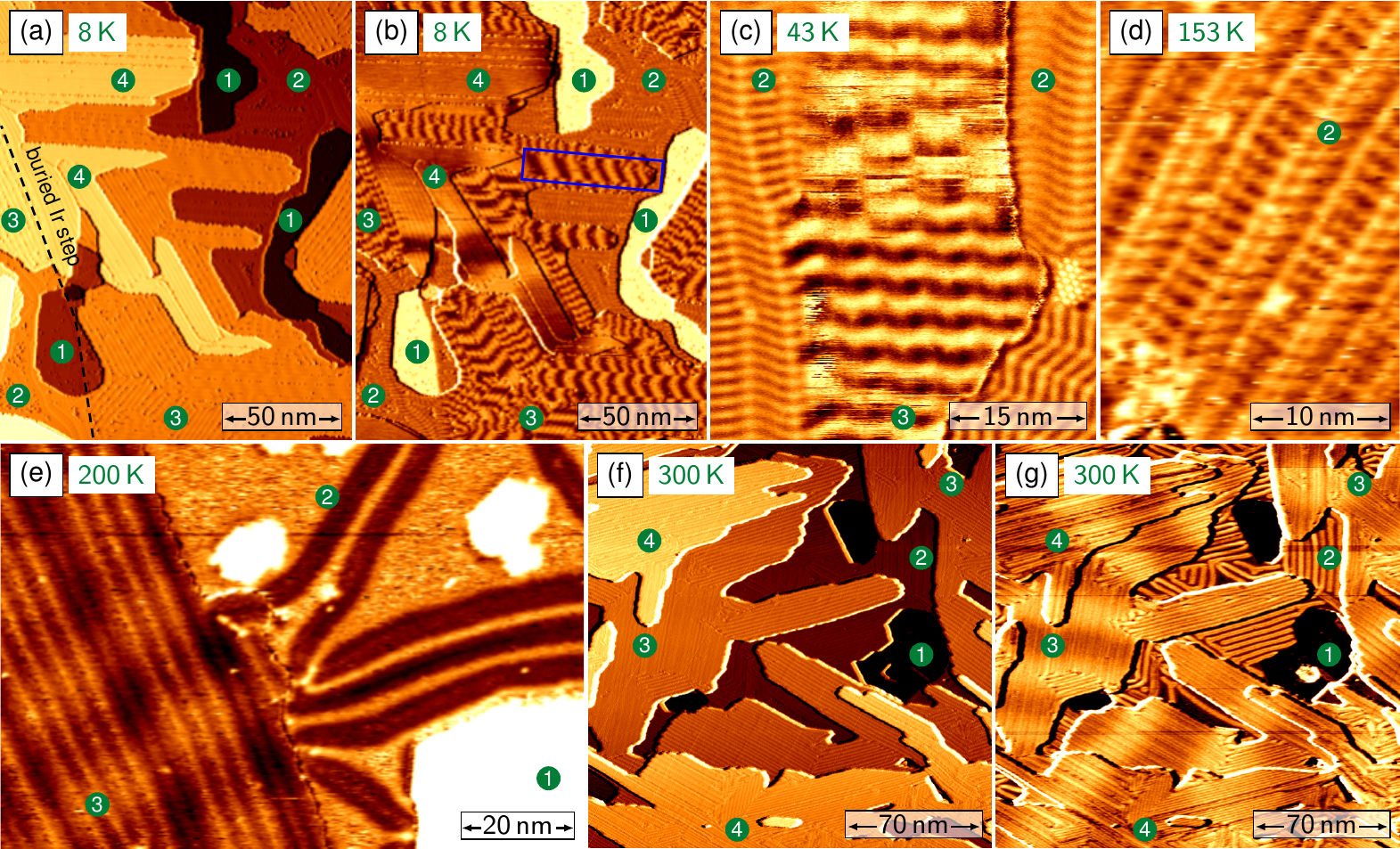}
  \caption{\label{TL_exp}  STM constant-current topography (a),(f) and
    spin-polarized differential conductance maps (b)-(e),(g) of ultrathin Fe
    films on Ir(111) for double-, triple-, and quadruple-layer coverages
    (indicated by the green circles).
    The simultaneously recorded current map was added to the constant-current
    topography in (a) and (f) to enhance the visibility of the dislocation
    lines. The spin spirals in the double layer are not visible at
    \SI{200}{\kelvin} anymore in (e), and the contrast visible in (g) in the
    double-layer regions is created by the dislocation lines.
    The spin spiral period in the triple
    layer gradually increases from \SI{4}{\nano\meter} in (b) to
    \SI{65}{\nano\meter} in (g).
    The quadruple layer is ferromagnetic in (b), whereas it forms a continuous
    spin spiral state with the triple layer in (g).
    The measurements shown in (a)-(b) and (f)-(g) were performed with an
    out-of-plane-sensitive antiferromagnetic Cr bulk tip, the ones in (c)-(e)
    with a ferromagnetic Fe coated W tip.
    Measurement parameters:
    (a)-(b) ${U = \SI{-700}{\milli\volt}}, {I = \SI{1}{\nano\ampere}}$;
    (c) ${U = \SI{-700}{\milli\volt}}, {I = \SI{0.6}{\nano\ampere}}$;
    (d) ${U = \SI{-500}{\milli\volt}}, {I = \SI{2}{\nano\ampere}}$;
    (e) ${U = \SI{-700}{\milli\volt}}, {I = \SI{2}{\nano\ampere}}$;
    (f)-(g) ${U = \SI{-500 }{\milli\volt}}, {I = \SI{3}{\nano\ampere}}$.
    }
\end{figure*}

In the monolayer, the experiments did not reveal a modification of the ratio
between the atomic and magnetic periods with the increase in
temperature~\cite{sonntag_thermal_2014}, which is also the
case for other noncollinear spin states at the nanometer scale, for
example in the  monolayer Mn~\cite{sessi_temperature_2009} and the double layer
Fe on W(110)~\cite{von_bergmann_coverage-dependent_2006}. However, in several
bulk materials it has already been observed~\cite{izyumov_modulated_1984} that
the period of the noncollinear order within the incommensurate phase may depend on
the temperature. For example, the period gradually increases with temperature by
less than 10\% in USb$_{0.9}$Te$_{0.1}$~\cite{Rossat-Mignod}, while it decreases
by about 30\% in Dy and Er~\cite{izyumov_modulated_1984}. In some cases, the
change in the magnetic period is connected to a similar modification of the
atomic structure with temperature, as it was demonstrated in bulk
ZnCr$_{2}$Se$_{4}$~\cite{Akimitsu}.

In this paper, we report on SP-STM measurements performed on ultrathin
Fe films on Ir(111) at various temperatures. We demonstrate that
the critical temperature of the noncollinear order significantly rises
from \SI{28}{\kelvin} in the monolayer~\cite{sonntag_thermal_2014}
through 150-\SI{200}{\kelvin} in the double layer to above room temperature in
the triple layer. Furthermore, we show that the period of the cycloidal spin
spiral state gradually increases from \SI{4}{\nano\meter} at \SI{8}{\kelvin}
to about \SI{65}{\nano\meter} at room temperature in the triple layer. Although
an enhancement of the period with temperature is not
unprecedented~\cite{izyumov_modulated_1984},
its magnitude is remarkably large in the present system. We
attribute this trend to the different strain relief and hybridization effects
between the three atomic layers, and construct a classical spin model with
layer-dependent interaction parameters to explain the phenomenon. For
ultrathin films thicker than a monolayer, an effective model with a mapping onto
a single layer might be too simplistic to describe the influence of
temperature on the spin structure. This could be important e.\,g. in the
calculation of magnetic phase diagrams, especially for multilayer systems
with many different interfaces.

% Details about the DL and TL at low temperature
The SP-STM measurements performed at different temperatures are illustrated in
Fig.~\ref{TL_exp}. The experimental methods are discussed in the Supplemental
Material~\cite{supp}.
Because of the large lattice mismatch between the Fe layer and the Ir substrate,
the films exhibit dislocation lines for coverages above a
monolayer~\cite{hsu_guiding_2016, hsu_electric-field-driven_2017,
  finco_tailoring_2016} as visible in the constant-current topography
image in Fig.~\ref{TL_exp}(a). The noncollinear magnetic structure of the Fe film at
low temperature is visible in the spin-polarized differential conductance
map in Fig.~\ref{TL_exp}(b) showing the out-of-plane component of the sample
magnetization. The alternating bright and dark stripes correspond to cycloidal
spin spirals propagating along the dislocation lines in both the double- and
triple-layer areas.  The distance between these lines is locally varying,
because the strain relief is not uniform. The period of the spin
spirals is correlated with the strain variation, which induces
a spreading of the wavelengths between 1.5 and \SI{2}{\nano\meter} in the
double layer~\cite{hsu_guiding_2016}.
Furthermore, two types of dislocation lines coexist in the triple layer. In
one case, the spin spirals have a period of 3 to
\SI{4}{\nano\meter} and a zigzag-shaped wavefront (clearly visible in
Fig.~\ref{TL_exp}(c)), whereas on the other type of area the periods are
ranging from 5 to \SI{10}{\nano\meter}
and the wavefront is straight but canted with respect to the dislocation lines
(see the blue box in Fig.~\ref{TL_exp}(b)). These differences arise from the
local arrangement of the atoms in the Fe layer~\cite{finco_tailoring_2016}.
In addition to previous reports, the magnetic state of the quadruple layer is
also shown in the image. It is ferromagnetic at low temperature on the length
scale of the island size (about \SI{100}{\nano\meter}), with the
domains assuming several different magnetization directions, which we attribute
to a weak magnetic anisotropy.

The temperature was raised in several steps up to room temperature. Up to
\SI{150}{\kelvin}, the wavefront shape of the spin spirals is conserved
in the double layer and their period stays within the range observed at low
temperature, see Fig.~\ref{TL_exp}(c) and (d). A possible increase of the
wavelength with temperature is too small to be distinguished from the
strain-induced variations. The magnetic
contrast vanishes between 150 and \SI{200}{\kelvin} in the double layer
(cf. Figs.~\ref{TL_exp}(d)-(e)),
which demonstrates the enhancement of the thermal stability over the
monolayer case where the nanoskyrmion lattice disappears at
\SI{28}{\kelvin}~\cite{sonntag_thermal_2014}.
However, since our SP-STM measurements are time-averaged,
this vanishing could be caused by spin fluctuations within a shorter time
scale than the measurement and thus the actual critical temperature for the
double layer might be higher~\cite{hasselberg_thermal_2015}.

In the triple layer, the period of the spin spirals increases clearly
and at \SI{200}{\kelvin}, the spiral wavefronts are
not zigzag-shaped anymore, with a wavelength of \SI{14}{\nano\meter} in the
area shown in Fig.~\ref{TL_exp}(e). At room temperature, the magnetism looks
strikingly different from the low-temperature case as illustrated in
Fig.~\ref{TL_exp}(g). The spiral
wavelengths are much larger (between 55 and \SI{80}{\nano\meter}) and all the
wavefronts are straight and perpendicular to the dislocation lines. Note that
the spirals are nevertheless still guided by the dislocation lines, but the
local atomic arrangement does not influence the wavefront anymore for such
large magnetic length scales. Furthermore, the spirals cross the step edges
between the triple- and quadruple-layer Fe areas, highlighting a ferromagnetic
coupling between the third and the fourth atomic layers which exhibit the same
magnetic wavelength.

\begin{figure}
  \includegraphics{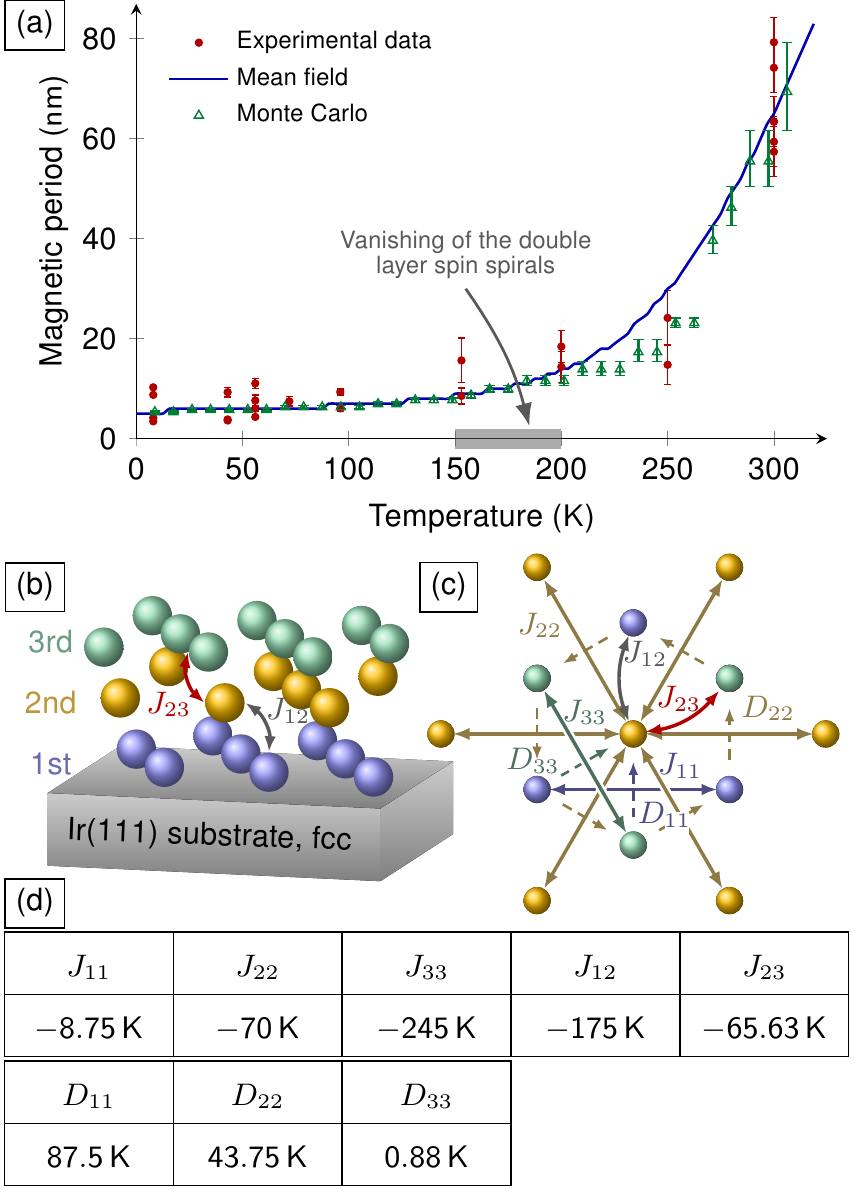}
  \caption{\label{plot} (a) Magnetic period of the spin spirals in the triple
    layer Fe on Ir(111)  at different temperatures from experiments, mean-field
    calculations and Monte Carlo simulations. The temperature scale for the
    mean-field data was rescaled by a factor of 0.71 for better comparison,
    since it significantly overestimates the temperature range of the period
    change as well as the critical temperature. The error bars reflect the
    resolution of the Fourier transformation (used to measure the period from
    the experimental data and from the simulations) as well as the thermal drift
    in the experiment.
    (b)-(c) Schematic view of the model system used
    to describe the triple-layer Fe on Ir(111) in the theoretical calculations,
    with the coupling constants used in Eq.~(\ref{eqn1}). Perfect fcc stacking
    was used to simplify the geometry of the model. The intralayer couplings
    are different in every layer, with
    $\left|J_{11}\right| < \left|J_{22}\right|< \left|J_{33}\right|$ for the
    Heisenberg couplings and $D_{11} > D_{22} > D_{33}$ for the
    Dzyaloshinsky-Moriya interactions.
    (d) Numerical values of the coupling constants used for the
    Monte Carlo simulations in (a), with the notations of (b)-(c).
 }
\end{figure}

In order to quantify the period increase in the triple layer, further
measurements were performed at intermediate temperatures and the data are
collected in Fig.~\ref{plot}(a).
On average, the period rises by a factor of 4 between \SI{8}{\kelvin}
and \SI{250}{\kelvin}, and again by a factor of 4 between \SI{250}{\kelvin} and
room temperature. The large dispersion of the data points at each temperature is
linked to the strain-relief effect discussed in Ref.~\cite{finco_tailoring_2016}.

%
% theory part
For a quantitative theoretical explanation of the remarkable increase of
the spiral wavelength with temperature, we relied on a model which treats
the three atomic layers of Fe separately instead of using single Heisenberg
and Dzyaloshinsky--Moriya interactions for the whole ultrathin film.
The model Hamiltonian reads
% In order to reproduce the increase of the spiral wavelength with the
% temperature in a theoretical model, an effective Hamiltonian with only
% nearest-neighbor interactions in a single layer is not sufficient and
% therefore we use the model Hamiltonian
\begin{eqnarray}
  \!\!\!\!\!\!H\!=\frac{1}{2}\!\!\!\sum_{p,q,\left<i,j\right>}\!\!\!J_{pq,ij}
  \boldsymbol{S}_{p,i}\boldsymbol{S}_{q,j}\!+\!\frac{1}{2}\!\!\!
  \sum_{p,\left<i,j\right>}\!\!\!\boldsymbol{D}_{pp,ij}
  \left(\boldsymbol{S}_{p,i}\times\boldsymbol{S}_{p,j}\right),\label{eqn1}
\end{eqnarray}
also illustrated in Figs.~\ref{plot}(b)-(c). The $\boldsymbol{S}_{p,i}$ denote
classical unit vectors representing the magnetic moments. The $p,q=1,2,3$
indices denote the three layers starting from the one closest to the Ir
substrate, while $i$ and $j$ are intralayer indices. The summations only
run over the nearest neighbors, including six intralayer neighbors and three
neighbors in each adjacent layer.

The $J_{pq,ij}$ coefficients denote intralayer and interlayer Heisenberg
exchange interactions. The intralayer Dzyaloshinsky--Moriya vectors
$\boldsymbol{D}_{pp,ij}$ are perpendicular to the nearest-neighbor bonds and
are chosen to be in-plane -- see the dashed arrows in Fig.~\ref{plot}(c).

Regarding the temperature-dependent spin spiral period, the crucial point about
Eq.~(\ref{eqn1}) is the consideration of layer-dependent interaction
parameters.
Layer-dependent coupling coefficients naturally appear during \textit{ab initio}
calculations~\cite{zakeri_direct_2013, meyerheim_new_2009,
  meng_direct_2014, dupe_engineering_2016} due to the different hybridization
effects for atomic layers
with different distances to the non-magnetic substrate and the vacuum interface.
In the present system, this choice is further supported by the fact
that the actual intralayer atomic distances may differ between the three Fe
layers due to the strain relief~\cite{hsu_electric-field-driven_2017,
  finco_tailoring_2016}. Note that the microscopic parameters in
Eq.~(\ref{eqn1}) are not explicitly temperature dependent. In an atomistic model
this is generally justified for Fe which has strong localized
magnetic moments~\cite{mryasov_temperature-dependent_2005}. In principle, a
modification of the
atomic structure with temperature could influence the coupling
coefficients~\cite{Akimitsu, veber_high-field_2008}. However, we did not observe
any obvious modification of the structure of the Fe film as a function of
temperature in the topographic images. In addition, the mechanisms discussed
in Ref.~\cite{izyumov_modulated_1984} for homogeneous bulk
magnets seemed to be insufficient to quantitatively explain the
exceptionally large increase of the period in the triple layer Fe.
% , although changes of the interlayer
%distances cannot be excluded.

For obtaining such a strong temperature dependence of the period, it was
necessary to assume that the $J_{pp}$ and $\boldsymbol{D}_{pp}$ intralayer
couplings determine different spiral periods in the different layers. The
Dzyaloshinsky--Moriya interaction is expected to get weaker when moving
away from the Ir substrate, since it primarily appears at the interface
between the magnetic Fe layers and the heavy metal Ir substrate with strong
spin--orbit coupling. On the contrary, the Heisenberg exchange interactions
should get stronger when moving away from the substrate, partly
because the rearrangement of the Fe atoms into the bcc
structure~\cite{hsu_electric-field-driven_2017} should make their values
approximate the strong ferromagnetic coupling in bcc Fe, and partly because
\textit{ab initio} calculations indicate a weakening of the Heisenberg
exchange interactions at Fe/Ir interfaces due to hybridization
effects~\cite{von_bergmann_observation_2006, simon_spin-correlations_2014,
  dupe_tailoring_2014, rozsa_skyrmions_2016}. As a net effect, the
determined period will be  higher in layers further away from the Ir
substrate.  It is also assumed that if only the intralayer couplings are
considered, then the critical temperature of the layers should increase
from the first layer towards the third.
% ; this
% is in agreement with the enhanced thermal stability with increasing coverage.
The interlayer ferromagnetic couplings $J_{12},J_{23}$ contribute to the
enhancement of the thermal stability, while ensuring that the layers are
coupled sufficiently strongly to each other, meaning that the period of
the spin spiral will be the same between the three layers at any fixed
temperature, and that the triple layer will only have a single critical
temperature.%  The experimental results do not indicate that the magnetic
% ordering should differ between the layers, in contrast to thin Fe films on
% Cu(001)~\cite{meyerheim_new_2009}.

How the period depends on the temperature can be understood from a
mean-field model. Following the derivation given in the Supplemental
Material~\cite{supp}, the free energy $F_{\textrm{MF}}$ per spin of the spin
spiral state with wave vector $k$ may be expressed as
\begin{eqnarray}
  \frac{1}{N}F_{\textrm{MF}}\left(k\right)=
  -\frac{1}{2}\sum_{p,q}\mathcal{J}_{pq}
     \left(k\right)\left<S_{p}\left(k\right)\right>
     \left<S_{q}\left(k\right)\right>\nonumber
\\
  -\sum_{p}k_{\textrm{B}}T\ln\left(4\pi\sinh\left(
     \frac{B_{p}\left(k\right)}{k_{\textrm{B}}T}\right)
     \frac{k_{\textrm{B}}T}{B_{p}\left(k\right)}\right),\label{eqn2}
\end{eqnarray}
with
\begin{eqnarray}
  B_{p}\left(k\right)=&&-\Bigg[\frac{1}{N}\sum_{q,\left<i,j\right>}J_{pq,ij}
                         \cos\left(k\left(x_{p,i}-x_{q,j}\right)\right)
                         \left<S_{q}\left(k\right)\right>\nonumber
\\
                      &&+D_{pq,ij}\sin\left(k\left(x_{p,i}-x_{q,j}\right)\right)
                         \left<S_{q}\left(k\right)\right>\Bigg]\label{eqn3}
\end{eqnarray}
the mean field in energy dimensions in layer $p$,
\begin{eqnarray}
  \mathcal{J}_{pq}\left(k\right)=
  &&\frac{1}{N}\sum_{\left<i,j\right>}J_{pq,ij}
     \cos\left(k\left(x_{p,i}-x_{q,j}\right)\right)\nonumber
\\
&&+D_{pq,ij}\sin\left(k\left(x_{p,i}-x_{q,j}\right)\right)\label{eqn4}
\end{eqnarray}
the Fourier transform of the interaction coefficients, and
$\left<S_{p}\left(k\right)\right>$ the order parameter of the spin spiral state.
$N$ denotes the number of atoms in a single layer. The equilibrium period of the
spin spiral may be obtained by minimizing Eq.~(\ref{eqn2}) with respect to the
wave vector $k$. The wave vector dependence of the free energy is included in
the coupling coefficients $\mathcal{J}_{pq}\left(k\right)$, which are minimized
by different $k$ values in the different layers as mentioned above. The
temperature dependence is encapsulated in the order parameters
$\left<S_{p}\left(k\right)\right>$, which do not depend significantly on the
wave vector. However,
$\left<S_{p}\left(k\right)\right>$ decreases faster with temperature in the
first and second layers than in the third one, gradually decreasing their
relative contribution to the free energy expression (\ref{eqn2}). This effect
shifts the minimum of $F_{\textrm{MF}}\left(k\right)$ towards lower wave vectors
with increasing temperature, explaining the effect observed in the experiments.

The interaction coefficients in Eq.~(\ref{eqn1}) were determined based on the
above assumptions regarding their relative magnitudes, and the numerical values
summarized in Fig.~\ref{plot}(d) were obtained by tuning the values in order to
quantitatively reproduce the temperature dependence observed in the experiments
by mean-field calculations and Monte Carlo simulations. Figure~\ref{plot}(c)
demonstrates a good agreement between the theoretical model and the
experiments. The details of the simulations are given in the Supplemental
Material~\cite{supp}.
Note that explaining the zigzag wavefront of the spin spirals~\cite{Hagemeister}
or the range of possible periods at low temperature~\cite{finco_tailoring_2016}
requires assumptions going beyond the model in Eq.~(\ref{eqn1}), but the
Hamiltonian seems to capture the main mechanism behind the temperature
dependence of the wavelength.
%
% Conclusion

In summary, we have shown using SP-STM measurements that
the observed magnetic period in the triple-layer Fe on Ir(111) increases
significantly, by approximately a factor of 16 between \SI{8}{\kelvin}
and \SI{300}{\kelvin}. Based on the different hybridization and strain relief
effects in the three atomic layers, we proposed a theoretical model with
layer-dependent coupling coefficients, which quantitatively reproduces the
period increase observed in the experiments. Since the presence of different
and coupled magnetic layers appears to be decisive, our work shows that the
usual mapping of ultrathin films onto a single effective layer might fail to
describe temperature effects or phase diagrams for films thicker than a single
atomic layer. Our results can hence motivate further investigations in magnetic
multilayers regarding the finite temperature behavior of noncollinear spin
structures.

\begin{acknowledgments}
  The authors thank T. Eelbo for technical help. Financial support by the
  European Union via the Horizon 2020 research and innovation programme under
  grant agreement No.~665095 (MagicSky), by the Deutsche Forschungsgemeinschaft
  via SFB668-A8 and -A11, by the Hamburgische Stiftung f\"{u}r Wissenschaften,
  Entwicklung und Kultur Helmut und Hannelore Greve, and by the National Research,
  Development and Innovation Office of Hungary under project No.~K115575 is
  gratefully acknowledged.
\end{acknowledgments}


\begin{thebibliography}{37}%
\makeatletter
\providecommand \@ifxundefined [1]{%
 \@ifx{#1\undefined}
}%
\providecommand \@ifnum [1]{%
 \ifnum #1\expandafter \@firstoftwo
 \else \expandafter \@secondoftwo
 \fi
}%
\providecommand \@ifx [1]{%
 \ifx #1\expandafter \@firstoftwo
 \else \expandafter \@secondoftwo
 \fi
}%
\providecommand \natexlab [1]{#1}%
\providecommand \enquote  [1]{``#1''}%
\providecommand \bibnamefont  [1]{#1}%
\providecommand \bibfnamefont [1]{#1}%
\providecommand \citenamefont [1]{#1}%
\providecommand \href@noop [0]{\@secondoftwo}%
\providecommand \href [0]{\begingroup \@sanitize@url \@href}%
\providecommand \@href[1]{\@@startlink{#1}\@@href}%
\providecommand \@@href[1]{\endgroup#1\@@endlink}%
\providecommand \@sanitize@url [0]{\catcode `\\12\catcode `\$12\catcode
  `\&12\catcode `\#12\catcode `\^12\catcode `\_12\catcode `\%12\relax}%
\providecommand \@@startlink[1]{}%
\providecommand \@@endlink[0]{}%
\providecommand \url  [0]{\begingroup\@sanitize@url \@url }%
\providecommand \@url [1]{\endgroup\@href {#1}{\urlprefix }}%
\providecommand \urlprefix  [0]{URL }%
\providecommand \Eprint [0]{\href }%
\providecommand \doibase [0]{http://dx.doi.org/}%
\providecommand \selectlanguage [0]{\@gobble}%
\providecommand \bibinfo  [0]{\@secondoftwo}%
\providecommand \bibfield  [0]{\@secondoftwo}%
\providecommand \translation [1]{[#1]}%
\providecommand \BibitemOpen [0]{}%
\providecommand \bibitemStop [0]{}%
\providecommand \bibitemNoStop [0]{.\EOS\space}%
\providecommand \EOS [0]{\spacefactor3000\relax}%
\providecommand \BibitemShut  [1]{\csname bibitem#1\endcsname}%
\let\auto@bib@innerbib\@empty
%</preamble>
\bibitem [{\citenamefont {Parkin}\ \emph {et~al.}(2008)\citenamefont {Parkin},
  \citenamefont {Hayashi},\ and\ \citenamefont {Thomas}}]{Parkin}%
  \BibitemOpen
  \bibfield  {author} {\bibinfo {author} {\bibfnamefont {S.~S.~P.}\
  \bibnamefont {Parkin}}, \bibinfo {author} {\bibfnamefont {M.}~\bibnamefont
  {Hayashi}}, \ and\ \bibinfo {author} {\bibfnamefont {L.}~\bibnamefont
  {Thomas}},\ }\href {\doibase 10.1126/science.1145799} {\bibfield  {journal}
  {\bibinfo  {journal} {Science}\ }\textbf {\bibinfo {volume} {320}},\ \bibinfo
  {pages} {190} (\bibinfo {year} {2008})}\BibitemShut {NoStop}%
\bibitem [{\citenamefont {Fert}\ \emph {et~al.}(2013)\citenamefont {Fert},
  \citenamefont {Cros},\ and\ \citenamefont {Sampaio}}]{Fert}%
  \BibitemOpen
  \bibfield  {author} {\bibinfo {author} {\bibfnamefont {A.}~\bibnamefont
  {Fert}}, \bibinfo {author} {\bibfnamefont {V.}~\bibnamefont {Cros}}, \ and\
  \bibinfo {author} {\bibfnamefont {J.}~\bibnamefont {Sampaio}},\ }\href@noop
  {} {\bibfield  {journal} {\bibinfo  {journal} {Nat. Nanotechnol.}\ }\textbf
  {\bibinfo {volume} {8}},\ \bibinfo {pages} {152} (\bibinfo {year}
  {2013})}\BibitemShut {NoStop}%
\bibitem [{\citenamefont {Dzyaloshinsky}(1958)}]{Dzyaloshinsky}%
  \BibitemOpen
  \bibfield  {author} {\bibinfo {author} {\bibfnamefont {I.}~\bibnamefont
  {Dzyaloshinsky}},\ }\href {\doibase
  http://dx.doi.org/10.1016/0022-3697(58)90076-3} {\bibfield  {journal}
  {\bibinfo  {journal} {J. Phys. Chem. Sol.}\ }\textbf {\bibinfo {volume}
  {4}},\ \bibinfo {pages} {241 } (\bibinfo {year} {1958})}\BibitemShut
  {NoStop}%
\bibitem [{\citenamefont {Moriya}(1960)}]{Moriya}%
  \BibitemOpen
  \bibfield  {author} {\bibinfo {author} {\bibfnamefont {T.}~\bibnamefont
  {Moriya}},\ }\href {\doibase 10.1103/PhysRevLett.4.228} {\bibfield  {journal}
  {\bibinfo  {journal} {Phys. Rev. Lett.}\ }\textbf {\bibinfo {volume} {4}},\
  \bibinfo {pages} {228} (\bibinfo {year} {1960})}\BibitemShut {NoStop}%
\bibitem [{\citenamefont {Moreau-Luchaire}\ \emph {et~al.}(2016)\citenamefont
  {Moreau-Luchaire}, \citenamefont {Moutaﬁs}, \citenamefont {Reyren},
  \citenamefont {Sampaio}, \citenamefont {Vaz}, \citenamefont {Horne},
  \citenamefont {Bouzehouane}, \citenamefont {Garcia}, \citenamefont
  {Deranlot}, \citenamefont {Warnicke}, \citenamefont {Wohlh\"{u}ter},
  \citenamefont {George}, \citenamefont {Weigand}, \citenamefont {Raabe},
  \citenamefont {Cros},\ and\ \citenamefont
  {Fert}}]{moreau-luchaire_additive_2016}%
  \BibitemOpen
  \bibfield  {author} {\bibinfo {author} {\bibfnamefont {C.}~\bibnamefont
  {Moreau-Luchaire}}, \bibinfo {author} {\bibfnamefont {C.}~\bibnamefont
  {Moutaﬁs}}, \bibinfo {author} {\bibfnamefont {N.}~\bibnamefont {Reyren}},
  \bibinfo {author} {\bibfnamefont {J.}~\bibnamefont {Sampaio}}, \bibinfo
  {author} {\bibfnamefont {C.~a.~F.}\ \bibnamefont {Vaz}}, \bibinfo {author}
  {\bibfnamefont {N.~V.}\ \bibnamefont {Horne}}, \bibinfo {author}
  {\bibfnamefont {K.}~\bibnamefont {Bouzehouane}}, \bibinfo {author}
  {\bibfnamefont {K.}~\bibnamefont {Garcia}}, \bibinfo {author} {\bibfnamefont
  {C.}~\bibnamefont {Deranlot}}, \bibinfo {author} {\bibfnamefont
  {P.}~\bibnamefont {Warnicke}}, \bibinfo {author} {\bibfnamefont
  {P.}~\bibnamefont {Wohlh\"{u}ter}}, \bibinfo {author} {\bibfnamefont {J.-M.}\
  \bibnamefont {George}}, \bibinfo {author} {\bibfnamefont {M.}~\bibnamefont
  {Weigand}}, \bibinfo {author} {\bibfnamefont {J.}~\bibnamefont {Raabe}},
  \bibinfo {author} {\bibfnamefont {V.}~\bibnamefont {Cros}}, \ and\ \bibinfo
  {author} {\bibfnamefont {A.}~\bibnamefont {Fert}},\ }\href {\doibase
  10.1038/nnano.2015.313} {\bibfield  {journal} {\bibinfo  {journal} {Nat.
  Nanotechnol.}\ }\textbf {\bibinfo {volume} {11}},\ \bibinfo {pages} {444}
  (\bibinfo {year} {2016})}\BibitemShut {NoStop}%
\bibitem [{\citenamefont {Chen}\ \emph {et~al.}(2015)\citenamefont {Chen},
  \citenamefont {Mascaraque}, \citenamefont {N'Diaye},\ and\ \citenamefont
  {Schmid}}]{chen_room_2015}%
  \BibitemOpen
  \bibfield  {author} {\bibinfo {author} {\bibfnamefont {G.}~\bibnamefont
  {Chen}}, \bibinfo {author} {\bibfnamefont {A.}~\bibnamefont {Mascaraque}},
  \bibinfo {author} {\bibfnamefont {A.~T.}\ \bibnamefont {N'Diaye}}, \ and\
  \bibinfo {author} {\bibfnamefont {A.~K.}\ \bibnamefont {Schmid}},\ }\href
  {\doibase 10.1063/1.4922726} {\bibfield  {journal} {\bibinfo  {journal}
  {Appl. Phys. Lett.}\ }\textbf {\bibinfo {volume} {106}},\ \bibinfo {pages}
  {242404} (\bibinfo {year} {2015})}\BibitemShut {NoStop}%
\bibitem [{\citenamefont {Jiang}\ \emph {et~al.}(2015)\citenamefont {Jiang},
  \citenamefont {Upadhyaya}, \citenamefont {Zhang}, \citenamefont {Yu},
  \citenamefont {Jungfleisch}, \citenamefont {Fradin}, \citenamefont {Pearson},
  \citenamefont {Tserkovnyak}, \citenamefont {Wang}, \citenamefont {Heinonen},
  \citenamefont {Velthuis},\ and\ \citenamefont
  {Hoffmann}}]{jiang_blowing_2015}%
  \BibitemOpen
  \bibfield  {author} {\bibinfo {author} {\bibfnamefont {W.}~\bibnamefont
  {Jiang}}, \bibinfo {author} {\bibfnamefont {P.}~\bibnamefont {Upadhyaya}},
  \bibinfo {author} {\bibfnamefont {W.}~\bibnamefont {Zhang}}, \bibinfo
  {author} {\bibfnamefont {G.}~\bibnamefont {Yu}}, \bibinfo {author}
  {\bibfnamefont {M.~B.}\ \bibnamefont {Jungfleisch}}, \bibinfo {author}
  {\bibfnamefont {F.~Y.}\ \bibnamefont {Fradin}}, \bibinfo {author}
  {\bibfnamefont {J.~E.}\ \bibnamefont {Pearson}}, \bibinfo {author}
  {\bibfnamefont {Y.}~\bibnamefont {Tserkovnyak}}, \bibinfo {author}
  {\bibfnamefont {K.~L.}\ \bibnamefont {Wang}}, \bibinfo {author}
  {\bibfnamefont {O.}~\bibnamefont {Heinonen}}, \bibinfo {author}
  {\bibfnamefont {S.~G. E.~t.}\ \bibnamefont {Velthuis}}, \ and\ \bibinfo
  {author} {\bibfnamefont {A.}~\bibnamefont {Hoffmann}},\ }\href {\doibase
  10.1126/science.aaa1442} {\bibfield  {journal} {\bibinfo  {journal}
  {Science}\ }\textbf {\bibinfo {volume} {349}},\ \bibinfo {pages} {283}
  (\bibinfo {year} {2015})}\BibitemShut {NoStop}%
\bibitem [{\citenamefont {Boulle}\ \emph {et~al.}(2016)\citenamefont {Boulle},
  \citenamefont {Vogel}, \citenamefont {Yang}, \citenamefont {Pizzini},
  \citenamefont {de~Souza~Chaves}, \citenamefont {Locatelli}, \citenamefont
  {Menteş}, \citenamefont {Sala}, \citenamefont {Buda-Prejbeanu},
  \citenamefont {Klein}, \citenamefont {Belmeguenai}, \citenamefont
  {Roussign\'{e}}, \citenamefont {Stashkevich}, \citenamefont {Ch\'{e}rif},
  \citenamefont {Aballe}, \citenamefont {Foerster}, \citenamefont {Chshiev},
  \citenamefont {Auffret}, \citenamefont {Miron},\ and\ \citenamefont
  {Gaudin}}]{boulle_room-temperature_2016}%
  \BibitemOpen
  \bibfield  {author} {\bibinfo {author} {\bibfnamefont {O.}~\bibnamefont
  {Boulle}}, \bibinfo {author} {\bibfnamefont {J.}~\bibnamefont {Vogel}},
  \bibinfo {author} {\bibfnamefont {H.}~\bibnamefont {Yang}}, \bibinfo {author}
  {\bibfnamefont {S.}~\bibnamefont {Pizzini}}, \bibinfo {author} {\bibfnamefont
  {D.}~\bibnamefont {de~Souza~Chaves}}, \bibinfo {author} {\bibfnamefont
  {A.}~\bibnamefont {Locatelli}}, \bibinfo {author} {\bibfnamefont {T.~O.}\
  \bibnamefont {Menteş}}, \bibinfo {author} {\bibfnamefont {A.}~\bibnamefont
  {Sala}}, \bibinfo {author} {\bibfnamefont {L.~D.}\ \bibnamefont
  {Buda-Prejbeanu}}, \bibinfo {author} {\bibfnamefont {O.}~\bibnamefont
  {Klein}}, \bibinfo {author} {\bibfnamefont {M.}~\bibnamefont {Belmeguenai}},
  \bibinfo {author} {\bibfnamefont {Y.}~\bibnamefont {Roussign\'{e}}}, \bibinfo
  {author} {\bibfnamefont {A.}~\bibnamefont {Stashkevich}}, \bibinfo {author}
  {\bibfnamefont {S.~M.}\ \bibnamefont {Ch\'{e}rif}}, \bibinfo {author}
  {\bibfnamefont {L.}~\bibnamefont {Aballe}}, \bibinfo {author} {\bibfnamefont
  {M.}~\bibnamefont {Foerster}}, \bibinfo {author} {\bibfnamefont
  {M.}~\bibnamefont {Chshiev}}, \bibinfo {author} {\bibfnamefont
  {S.}~\bibnamefont {Auffret}}, \bibinfo {author} {\bibfnamefont {I.~M.}\
  \bibnamefont {Miron}}, \ and\ \bibinfo {author} {\bibfnamefont
  {G.}~\bibnamefont {Gaudin}},\ }\href {\doibase 10.1038/nnano.2015.315}
  {\bibfield  {journal} {\bibinfo  {journal} {Nat. Nanotechnol.}\ }\textbf
  {\bibinfo {volume} {11}},\ \bibinfo {pages} {449} (\bibinfo {year}
  {2016})}\BibitemShut {NoStop}%
\bibitem [{\citenamefont {Heinze}\ \emph {et~al.}(2011)\citenamefont {Heinze},
  \citenamefont {von Bergmann}, \citenamefont {Menzel}, \citenamefont {Brede},
  \citenamefont {Kubetzka}, \citenamefont {Wiesendanger}, \citenamefont
  {Bihlmayer},\ and\ \citenamefont {Bl\"{u}gel}}]{heinze_spontaneous_2011}%
  \BibitemOpen
  \bibfield  {author} {\bibinfo {author} {\bibfnamefont {S.}~\bibnamefont
  {Heinze}}, \bibinfo {author} {\bibfnamefont {K.}~\bibnamefont {von
  Bergmann}}, \bibinfo {author} {\bibfnamefont {M.}~\bibnamefont {Menzel}},
  \bibinfo {author} {\bibfnamefont {J.}~\bibnamefont {Brede}}, \bibinfo
  {author} {\bibfnamefont {A.}~\bibnamefont {Kubetzka}}, \bibinfo {author}
  {\bibfnamefont {R.}~\bibnamefont {Wiesendanger}}, \bibinfo {author}
  {\bibfnamefont {G.}~\bibnamefont {Bihlmayer}}, \ and\ \bibinfo {author}
  {\bibfnamefont {S.}~\bibnamefont {Bl\"{u}gel}},\ }\href {\doibase
  10.1038/nphys2045} {\bibfield  {journal} {\bibinfo  {journal} {Nat. Phys.}\
  }\textbf {\bibinfo {volume} {7}},\ \bibinfo {pages} {713} (\bibinfo {year}
  {2011})}\BibitemShut {NoStop}%
\bibitem [{\citenamefont {Sonntag}\ \emph {et~al.}(2014)\citenamefont
  {Sonntag}, \citenamefont {Hermenau}, \citenamefont {Krause},\ and\
  \citenamefont {Wiesendanger}}]{sonntag_thermal_2014}%
  \BibitemOpen
  \bibfield  {author} {\bibinfo {author} {\bibfnamefont {A.}~\bibnamefont
  {Sonntag}}, \bibinfo {author} {\bibfnamefont {J.}~\bibnamefont {Hermenau}},
  \bibinfo {author} {\bibfnamefont {S.}~\bibnamefont {Krause}}, \ and\ \bibinfo
  {author} {\bibfnamefont {R.}~\bibnamefont {Wiesendanger}},\ }\href {\doibase
  10.1103/PhysRevLett.113.077202} {\bibfield  {journal} {\bibinfo  {journal}
  {Phys. Rev. Lett.}\ }\textbf {\bibinfo {volume} {113}},\ \bibinfo {pages}
  {077202} (\bibinfo {year} {2014})}\BibitemShut {NoStop}%
\bibitem [{\citenamefont {Jensen}\ \emph {et~al.}(1992)\citenamefont {Jensen},
  \citenamefont {Dreyss\'{e}},\ and\ \citenamefont
  {Bennemann}}]{jensen_calculation_1992}%
  \BibitemOpen
  \bibfield  {author} {\bibinfo {author} {\bibfnamefont {P.~J.}\ \bibnamefont
  {Jensen}}, \bibinfo {author} {\bibfnamefont {H.}~\bibnamefont {Dreyss\'{e}}},
  \ and\ \bibinfo {author} {\bibfnamefont {K.~H.}\ \bibnamefont {Bennemann}},\
  }\href {http://iopscience.iop.org/article/10.1209/0295-5075/18/5/015/meta}
  {\bibfield  {journal} {\bibinfo  {journal} {Europhys. Lett.}\ }\textbf
  {\bibinfo {volume} {18}},\ \bibinfo {pages} {463} (\bibinfo {year}
  {1992})}\BibitemShut {NoStop}%
\bibitem [{\citenamefont {Schneider}\ \emph {et~al.}(1990)\citenamefont
  {Schneider}, \citenamefont {Bressler}, \citenamefont {Schuster},
  \citenamefont {Kirschner}, \citenamefont {de~Miguel},\ and\ \citenamefont
  {Miranda}}]{schneider_curie_1990}%
  \BibitemOpen
  \bibfield  {author} {\bibinfo {author} {\bibfnamefont {C.~M.}\ \bibnamefont
  {Schneider}}, \bibinfo {author} {\bibfnamefont {P.}~\bibnamefont {Bressler}},
  \bibinfo {author} {\bibfnamefont {P.}~\bibnamefont {Schuster}}, \bibinfo
  {author} {\bibfnamefont {J.}~\bibnamefont {Kirschner}}, \bibinfo {author}
  {\bibfnamefont {J.~J.}\ \bibnamefont {de~Miguel}}, \ and\ \bibinfo {author}
  {\bibfnamefont {R.}~\bibnamefont {Miranda}},\ }\href {\doibase
  10.1103/PhysRevLett.64.1059} {\bibfield  {journal} {\bibinfo  {journal}
  {Phys. Rev. Lett.}\ }\textbf {\bibinfo {volume} {64}},\ \bibinfo {pages}
  {1059} (\bibinfo {year} {1990})}\BibitemShut {NoStop}%
\bibitem [{\citenamefont {Elmers}\ \emph {et~al.}(1995)\citenamefont {Elmers},
  \citenamefont {Hauschild}, \citenamefont {Fritzsche}, \citenamefont {Liu},
  \citenamefont {Gradmann},\ and\ \citenamefont
  {K\"{o}hler}}]{elmers_magnetic_1995}%
  \BibitemOpen
  \bibfield  {author} {\bibinfo {author} {\bibfnamefont {H.~J.}\ \bibnamefont
  {Elmers}}, \bibinfo {author} {\bibfnamefont {J.}~\bibnamefont {Hauschild}},
  \bibinfo {author} {\bibfnamefont {H.}~\bibnamefont {Fritzsche}}, \bibinfo
  {author} {\bibfnamefont {G.}~\bibnamefont {Liu}}, \bibinfo {author}
  {\bibfnamefont {U.}~\bibnamefont {Gradmann}}, \ and\ \bibinfo {author}
  {\bibfnamefont {U.}~\bibnamefont {K\"{o}hler}},\ }\href
  {https://journals.aps.org/prl/abstract/10.1103/PhysRevLett.75.2031}
  {\bibfield  {journal} {\bibinfo  {journal} {Physical Review Letters}\
  }\textbf {\bibinfo {volume} {75}},\ \bibinfo {pages} {2031} (\bibinfo {year}
  {1995})}\BibitemShut {NoStop}%
\bibitem [{\citenamefont {Sessi}\ \emph {et~al.}(2009)\citenamefont {Sessi},
  \citenamefont {Guisinger}, \citenamefont {Guest},\ and\ \citenamefont
  {Bode}}]{sessi_temperature_2009}%
  \BibitemOpen
  \bibfield  {author} {\bibinfo {author} {\bibfnamefont {P.}~\bibnamefont
  {Sessi}}, \bibinfo {author} {\bibfnamefont {N.~P.}\ \bibnamefont
  {Guisinger}}, \bibinfo {author} {\bibfnamefont {J.~R.}\ \bibnamefont
  {Guest}}, \ and\ \bibinfo {author} {\bibfnamefont {M.}~\bibnamefont {Bode}},\
  }\href {\doibase 10.1103/PhysRevLett.103.167201} {\bibfield  {journal}
  {\bibinfo  {journal} {Phys. Rev. Lett.}\ }\textbf {\bibinfo {volume} {103}},\
  \bibinfo {pages} {167201} (\bibinfo {year} {2009})}\BibitemShut {NoStop}%
\bibitem [{\citenamefont {von Bergmann}\ \emph
  {et~al.}(2006{\natexlab{a}})\citenamefont {von Bergmann}, \citenamefont
  {Bode},\ and\ \citenamefont
  {Wiesendanger}}]{von_bergmann_coverage-dependent_2006}%
  \BibitemOpen
  \bibfield  {author} {\bibinfo {author} {\bibfnamefont {K.}~\bibnamefont {von
  Bergmann}}, \bibinfo {author} {\bibfnamefont {M.}~\bibnamefont {Bode}}, \
  and\ \bibinfo {author} {\bibfnamefont {R.}~\bibnamefont {Wiesendanger}},\
  }\href {\doibase 10.1016/j.jmmm.2005.12.015} {\bibfield  {journal} {\bibinfo
  {journal} {J. Magn. Magn. Mater.}\ }\textbf {\bibinfo {volume} {305}},\
  \bibinfo {pages} {279} (\bibinfo {year} {2006}{\natexlab{a}})}\BibitemShut
  {NoStop}%
\bibitem [{\citenamefont {Izyumov}(1984)}]{izyumov_modulated_1984}%
  \BibitemOpen
  \bibfield  {author} {\bibinfo {author} {\bibfnamefont {Y.~A.}\ \bibnamefont
  {Izyumov}},\ }\href {\doibase 10.1070/PU1984v027n11ABEH004120} {\bibfield
  {journal} {\bibinfo  {journal} {Sov. Phys. Usp.}\ }\textbf {\bibinfo {volume}
  {27}},\ \bibinfo {pages} {845} (\bibinfo {year} {1984})}\BibitemShut
  {NoStop}%
\bibitem [{\citenamefont {Rossat-Mignod}\ \emph {et~al.}(1979)\citenamefont
  {Rossat-Mignod}, \citenamefont {Burlet}, \citenamefont {Vogt},\ and\
  \citenamefont {Lander}}]{Rossat-Mignod}%
  \BibitemOpen
  \bibfield  {author} {\bibinfo {author} {\bibfnamefont {J.}~\bibnamefont
  {Rossat-Mignod}}, \bibinfo {author} {\bibfnamefont {P.}~\bibnamefont
  {Burlet}}, \bibinfo {author} {\bibfnamefont {O.}~\bibnamefont {Vogt}}, \ and\
  \bibinfo {author} {\bibfnamefont {G.~H.}\ \bibnamefont {Lander}},\ }\href
  {http://stacks.iop.org/0022-3719/12/i=6/a=021} {\bibfield  {journal}
  {\bibinfo  {journal} {J. Phys. C: Sol. State Phys.}\ }\textbf {\bibinfo
  {volume} {12}},\ \bibinfo {pages} {1101} (\bibinfo {year}
  {1979})}\BibitemShut {NoStop}%
\bibitem [{\citenamefont {Akimitsu}\ \emph {et~al.}(1978)\citenamefont
  {Akimitsu}, \citenamefont {Siratori}, \citenamefont {Shirane}, \citenamefont
  {Iizumi},\ and\ \citenamefont {Watanabe}}]{Akimitsu}%
  \BibitemOpen
  \bibfield  {author} {\bibinfo {author} {\bibfnamefont {J.}~\bibnamefont
  {Akimitsu}}, \bibinfo {author} {\bibfnamefont {K.}~\bibnamefont {Siratori}},
  \bibinfo {author} {\bibfnamefont {G.}~\bibnamefont {Shirane}}, \bibinfo
  {author} {\bibfnamefont {M.}~\bibnamefont {Iizumi}}, \ and\ \bibinfo {author}
  {\bibfnamefont {T.}~\bibnamefont {Watanabe}},\ }\href {\doibase
  10.1143/JPSJ.44.172} {\bibfield  {journal} {\bibinfo  {journal} {J. Phys.
  Soc. Jpn.}\ }\textbf {\bibinfo {volume} {44}},\ \bibinfo {pages} {172}
  (\bibinfo {year} {1978})}\BibitemShut {NoStop}%
\bibitem [{sup()}]{supp}%
  \BibitemOpen
  \href@noop {} {\bibinfo  {journal} {Supplemental Material describing the
  details of the experiments, the mean-field calculations and the Monte Carlo
  simulations. It also includes
  Refs.~\cite{arblaster_crystallographic_2010,rozsa_magnetic_2015,rozsa_complex_2016}}\
  }\BibitemShut {NoStop}%
\bibitem [{\citenamefont {Hsu}\ \emph {et~al.}(2016)\citenamefont {Hsu},
  \citenamefont {Finco}, \citenamefont {Schmidt}, \citenamefont {Kubetzka},
  \citenamefont {von Bergmann},\ and\ \citenamefont
  {Wiesendanger}}]{hsu_guiding_2016}%
  \BibitemOpen
\bibfield  {journal} {  }\bibfield  {author} {\bibinfo {author} {\bibfnamefont
  {P.-J.}\ \bibnamefont {Hsu}}, \bibinfo {author} {\bibfnamefont
  {A.}~\bibnamefont {Finco}}, \bibinfo {author} {\bibfnamefont
  {L.}~\bibnamefont {Schmidt}}, \bibinfo {author} {\bibfnamefont
  {A.}~\bibnamefont {Kubetzka}}, \bibinfo {author} {\bibfnamefont
  {K.}~\bibnamefont {von Bergmann}}, \ and\ \bibinfo {author} {\bibfnamefont
  {R.}~\bibnamefont {Wiesendanger}},\ }\href {\doibase
  10.1103/PhysRevLett.116.017201} {\bibfield  {journal} {\bibinfo  {journal}
  {Phys. Rev. Lett.}\ }\textbf {\bibinfo {volume} {116}},\ \bibinfo {pages}
  {017201} (\bibinfo {year} {2016})}\BibitemShut {NoStop}%
\bibitem [{\citenamefont {Hsu}\ \emph {et~al.}(2017)\citenamefont {Hsu},
  \citenamefont {Kubetzka}, \citenamefont {Finco}, \citenamefont {Romming},
  \citenamefont {von Bergmann},\ and\ \citenamefont
  {Wiesendanger}}]{hsu_electric-field-driven_2017}%
  \BibitemOpen
  \bibfield  {author} {\bibinfo {author} {\bibfnamefont {P.-J.}\ \bibnamefont
  {Hsu}}, \bibinfo {author} {\bibfnamefont {A.}~\bibnamefont {Kubetzka}},
  \bibinfo {author} {\bibfnamefont {A.}~\bibnamefont {Finco}}, \bibinfo
  {author} {\bibfnamefont {N.}~\bibnamefont {Romming}}, \bibinfo {author}
  {\bibfnamefont {K.}~\bibnamefont {von Bergmann}}, \ and\ \bibinfo {author}
  {\bibfnamefont {R.}~\bibnamefont {Wiesendanger}},\ }\href {\doibase
  doi:10.1038/nnano.2016.234} {\bibfield  {journal} {\bibinfo  {journal} {Nat.
  Nanotechnol.}\ }\textbf {\bibinfo {volume} {12}},\ \bibinfo {pages} {123}
  (\bibinfo {year} {2017})}\BibitemShut {NoStop}%
\bibitem [{\citenamefont {Finco}\ \emph {et~al.}(2016)\citenamefont {Finco},
  \citenamefont {Hsu}, \citenamefont {Kubetzka}, \citenamefont {von Bergmann},\
  and\ \citenamefont {Wiesendanger}}]{finco_tailoring_2016}%
  \BibitemOpen
  \bibfield  {author} {\bibinfo {author} {\bibfnamefont {A.}~\bibnamefont
  {Finco}}, \bibinfo {author} {\bibfnamefont {P.-J.}\ \bibnamefont {Hsu}},
  \bibinfo {author} {\bibfnamefont {A.}~\bibnamefont {Kubetzka}}, \bibinfo
  {author} {\bibfnamefont {K.}~\bibnamefont {von Bergmann}}, \ and\ \bibinfo
  {author} {\bibfnamefont {R.}~\bibnamefont {Wiesendanger}},\ }\href {\doibase
  10.1103/PhysRevB.94.214402} {\bibfield  {journal} {\bibinfo  {journal} {Phys.
  Rev. B}\ }\textbf {\bibinfo {volume} {94}},\ \bibinfo {pages} {214402}
  (\bibinfo {year} {2016})}\BibitemShut {NoStop}%
\bibitem [{\citenamefont {Hasselberg}\ \emph {et~al.}(2015)\citenamefont
  {Hasselberg}, \citenamefont {Yanes}, \citenamefont {Hinzke}, \citenamefont
  {Sessi}, \citenamefont {Bode}, \citenamefont {Szunyogh},\ and\ \citenamefont
  {Nowak}}]{hasselberg_thermal_2015}%
  \BibitemOpen
  \bibfield  {author} {\bibinfo {author} {\bibfnamefont {G.}~\bibnamefont
  {Hasselberg}}, \bibinfo {author} {\bibfnamefont {R.}~\bibnamefont {Yanes}},
  \bibinfo {author} {\bibfnamefont {D.}~\bibnamefont {Hinzke}}, \bibinfo
  {author} {\bibfnamefont {P.}~\bibnamefont {Sessi}}, \bibinfo {author}
  {\bibfnamefont {M.}~\bibnamefont {Bode}}, \bibinfo {author} {\bibfnamefont
  {L.}~\bibnamefont {Szunyogh}}, \ and\ \bibinfo {author} {\bibfnamefont
  {U.}~\bibnamefont {Nowak}},\ }\href
  {http://link.aps.org/doi/10.1103/PhysRevB.91.064402} {\bibfield  {journal}
  {\bibinfo  {journal} {Phys. Rev. B}\ }\textbf {\bibinfo {volume} {91}},\
  \bibinfo {pages} {064402} (\bibinfo {year} {2015})}\BibitemShut {NoStop}%
\bibitem [{\citenamefont {Zakeri}\ \emph {et~al.}(2013)\citenamefont {Zakeri},
  \citenamefont {Chuang}, \citenamefont {Ernst}, \citenamefont {Sandratskii},
  \citenamefont {Buczek}, \citenamefont {Qin}, \citenamefont {Zhang},\ and\
  \citenamefont {Kirschner}}]{zakeri_direct_2013}%
  \BibitemOpen
  \bibfield  {author} {\bibinfo {author} {\bibfnamefont {K.}~\bibnamefont
  {Zakeri}}, \bibinfo {author} {\bibfnamefont {T.-H.}\ \bibnamefont {Chuang}},
  \bibinfo {author} {\bibfnamefont {A.}~\bibnamefont {Ernst}}, \bibinfo
  {author} {\bibfnamefont {L.~M.}\ \bibnamefont {Sandratskii}}, \bibinfo
  {author} {\bibfnamefont {P.}~\bibnamefont {Buczek}}, \bibinfo {author}
  {\bibfnamefont {H.~J.}\ \bibnamefont {Qin}}, \bibinfo {author} {\bibfnamefont
  {Y.}~\bibnamefont {Zhang}}, \ and\ \bibinfo {author} {\bibfnamefont
  {J.}~\bibnamefont {Kirschner}},\ }\href {\doibase 10.1038/nnano.2013.188}
  {\bibfield  {journal} {\bibinfo  {journal} {Nat. Nanotechnol.}\ }\textbf
  {\bibinfo {volume} {8}},\ \bibinfo {pages} {853} (\bibinfo {year}
  {2013})}\BibitemShut {NoStop}%
\bibitem [{\citenamefont {Meyerheim}\ \emph {et~al.}(2009)\citenamefont
  {Meyerheim}, \citenamefont {Tonnerre}, \citenamefont {Sandratskii},
  \citenamefont {Tolentino}, \citenamefont {Przybylski}, \citenamefont {Gabi},
  \citenamefont {Yildiz}, \citenamefont {Fu}, \citenamefont {Bontempi},
  \citenamefont {Grenier},\ and\ \citenamefont
  {Kirschner}}]{meyerheim_new_2009}%
  \BibitemOpen
  \bibfield  {author} {\bibinfo {author} {\bibfnamefont {H.~L.}\ \bibnamefont
  {Meyerheim}}, \bibinfo {author} {\bibfnamefont {J.-M.}\ \bibnamefont
  {Tonnerre}}, \bibinfo {author} {\bibfnamefont {L.}~\bibnamefont
  {Sandratskii}}, \bibinfo {author} {\bibfnamefont {H.~C.~N.}\ \bibnamefont
  {Tolentino}}, \bibinfo {author} {\bibfnamefont {M.}~\bibnamefont
  {Przybylski}}, \bibinfo {author} {\bibfnamefont {Y.}~\bibnamefont {Gabi}},
  \bibinfo {author} {\bibfnamefont {F.}~\bibnamefont {Yildiz}}, \bibinfo
  {author} {\bibfnamefont {X.~L.}\ \bibnamefont {Fu}}, \bibinfo {author}
  {\bibfnamefont {E.}~\bibnamefont {Bontempi}}, \bibinfo {author}
  {\bibfnamefont {S.}~\bibnamefont {Grenier}}, \ and\ \bibinfo {author}
  {\bibfnamefont {J.}~\bibnamefont {Kirschner}},\ }\href {\doibase
  10.1103/PhysRevLett.103.267202} {\bibfield  {journal} {\bibinfo  {journal}
  {Phys. Rev. Lett.}\ }\textbf {\bibinfo {volume} {103}},\ \bibinfo {pages}
  {267202} (\bibinfo {year} {2009})}\BibitemShut {NoStop}%
\bibitem [{\citenamefont {Meng}\ \emph {et~al.}(2014)\citenamefont {Meng},
  \citenamefont {Zakeri}, \citenamefont {Ernst}, \citenamefont {Chuang},
  \citenamefont {Qin}, \citenamefont {Chen},\ and\ \citenamefont
  {Kirschner}}]{meng_direct_2014}%
  \BibitemOpen
  \bibfield  {author} {\bibinfo {author} {\bibfnamefont {Y.}~\bibnamefont
  {Meng}}, \bibinfo {author} {\bibfnamefont {K.}~\bibnamefont {Zakeri}},
  \bibinfo {author} {\bibfnamefont {A.}~\bibnamefont {Ernst}}, \bibinfo
  {author} {\bibfnamefont {T.-H.}\ \bibnamefont {Chuang}}, \bibinfo {author}
  {\bibfnamefont {H.~J.}\ \bibnamefont {Qin}}, \bibinfo {author} {\bibfnamefont
  {Y.-J.}\ \bibnamefont {Chen}}, \ and\ \bibinfo {author} {\bibfnamefont
  {J.}~\bibnamefont {Kirschner}},\ }\href {\doibase 10.1103/PhysRevB.90.174437}
  {\bibfield  {journal} {\bibinfo  {journal} {Phys. Rev. B}\ }\textbf {\bibinfo
  {volume} {90}},\ \bibinfo {pages} {174437} (\bibinfo {year}
  {2014})}\BibitemShut {NoStop}%
\bibitem [{\citenamefont {Dup\'{e}}\ \emph {et~al.}(2016)\citenamefont
  {Dup\'{e}}, \citenamefont {Bihlmayer}, \citenamefont {B\"{o}ttcher},
  \citenamefont {Bl\"{u}gel},\ and\ \citenamefont
  {Heinze}}]{dupe_engineering_2016}%
  \BibitemOpen
  \bibfield  {author} {\bibinfo {author} {\bibfnamefont {B.}~\bibnamefont
  {Dup\'{e}}}, \bibinfo {author} {\bibfnamefont {G.}~\bibnamefont {Bihlmayer}},
  \bibinfo {author} {\bibfnamefont {M.}~\bibnamefont {B\"{o}ttcher}}, \bibinfo
  {author} {\bibfnamefont {S.}~\bibnamefont {Bl\"{u}gel}}, \ and\ \bibinfo
  {author} {\bibfnamefont {S.}~\bibnamefont {Heinze}},\ }\href {\doibase
  10.1038/ncomms11779} {\bibfield  {journal} {\bibinfo  {journal} {Nature
  Communications}\ }\textbf {\bibinfo {volume} {7}},\ \bibinfo {pages} {11779}
  (\bibinfo {year} {2016})}\BibitemShut {NoStop}%
\bibitem [{\citenamefont {Mryasov}\ \emph {et~al.}(2005)\citenamefont
  {Mryasov}, \citenamefont {Nowak}, \citenamefont {Guslienko},\ and\
  \citenamefont {Chantrell}}]{mryasov_temperature-dependent_2005}%
  \BibitemOpen
  \bibfield  {author} {\bibinfo {author} {\bibfnamefont {O.~N.}\ \bibnamefont
  {Mryasov}}, \bibinfo {author} {\bibfnamefont {U.}~\bibnamefont {Nowak}},
  \bibinfo {author} {\bibfnamefont {K.~Y.}\ \bibnamefont {Guslienko}}, \ and\
  \bibinfo {author} {\bibfnamefont {R.~W.}\ \bibnamefont {Chantrell}},\ }\href
  {\doibase 10.1209/epl/i2004-10404-2} {\bibfield  {journal} {\bibinfo
  {journal} {Europhysics Letters (EPL)}\ }\textbf {\bibinfo {volume} {69}},\
  \bibinfo {pages} {805} (\bibinfo {year} {2005})}\BibitemShut {NoStop}%
\bibitem [{\citenamefont {Veber}\ \emph {et~al.}(2008)\citenamefont {Veber},
  \citenamefont {Fedin}, \citenamefont {Potapov}, \citenamefont {Maryunina},
  \citenamefont {Romanenko}, \citenamefont {Sagdeev}, \citenamefont
  {Ovcharenko}, \citenamefont {Goldfarb},\ and\ \citenamefont
  {Bagryanskaya}}]{veber_high-field_2008}%
  \BibitemOpen
  \bibfield  {author} {\bibinfo {author} {\bibfnamefont {S.~L.}\ \bibnamefont
  {Veber}}, \bibinfo {author} {\bibfnamefont {M.~V.}\ \bibnamefont {Fedin}},
  \bibinfo {author} {\bibfnamefont {A.~I.}\ \bibnamefont {Potapov}}, \bibinfo
  {author} {\bibfnamefont {K.~Y.}\ \bibnamefont {Maryunina}}, \bibinfo {author}
  {\bibfnamefont {G.~V.}\ \bibnamefont {Romanenko}}, \bibinfo {author}
  {\bibfnamefont {R.~Z.}\ \bibnamefont {Sagdeev}}, \bibinfo {author}
  {\bibfnamefont {V.~I.}\ \bibnamefont {Ovcharenko}}, \bibinfo {author}
  {\bibfnamefont {D.}~\bibnamefont {Goldfarb}}, \ and\ \bibinfo {author}
  {\bibfnamefont {E.~G.}\ \bibnamefont {Bagryanskaya}},\ }\href {\doibase
  10.1021/ja710773u} {\bibfield  {journal} {\bibinfo  {journal} {Journal of the
  American Chemical Society}\ }\textbf {\bibinfo {volume} {130}},\ \bibinfo
  {pages} {2444} (\bibinfo {year} {2008})}\BibitemShut {NoStop}%
\bibitem [{\citenamefont {von Bergmann}\ \emph
  {et~al.}(2006{\natexlab{b}})\citenamefont {von Bergmann}, \citenamefont
  {Heinze}, \citenamefont {Bode}, \citenamefont {Vedmedenko}, \citenamefont
  {Bihlmayer}, \citenamefont {Bl\"{u}gel},\ and\ \citenamefont
  {Wiesendanger}}]{von_bergmann_observation_2006}%
  \BibitemOpen
  \bibfield  {author} {\bibinfo {author} {\bibfnamefont {K.}~\bibnamefont {von
  Bergmann}}, \bibinfo {author} {\bibfnamefont {S.}~\bibnamefont {Heinze}},
  \bibinfo {author} {\bibfnamefont {M.}~\bibnamefont {Bode}}, \bibinfo {author}
  {\bibfnamefont {E.~Y.}\ \bibnamefont {Vedmedenko}}, \bibinfo {author}
  {\bibfnamefont {G.}~\bibnamefont {Bihlmayer}}, \bibinfo {author}
  {\bibfnamefont {S.}~\bibnamefont {Bl\"{u}gel}}, \ and\ \bibinfo {author}
  {\bibfnamefont {R.}~\bibnamefont {Wiesendanger}},\ }\href {\doibase
  10.1103/PhysRevLett.96.167203} {\bibfield  {journal} {\bibinfo  {journal}
  {Physical Review Letters}\ }\textbf {\bibinfo {volume} {96}},\ \bibinfo
  {pages} {167203} (\bibinfo {year} {2006}{\natexlab{b}})}\BibitemShut
  {NoStop}%
\bibitem [{\citenamefont {Simon}\ \emph {et~al.}(2014)\citenamefont {Simon},
  \citenamefont {Palot\'{a}s}, \citenamefont {Ujfalussy}, \citenamefont
  {De\'{a}k}, \citenamefont {Stocks},\ and\ \citenamefont
  {Szunyogh}}]{simon_spin-correlations_2014}%
  \BibitemOpen
  \bibfield  {author} {\bibinfo {author} {\bibfnamefont {E.}~\bibnamefont
  {Simon}}, \bibinfo {author} {\bibfnamefont {K.}~\bibnamefont {Palot\'{a}s}},
  \bibinfo {author} {\bibfnamefont {B.}~\bibnamefont {Ujfalussy}}, \bibinfo
  {author} {\bibfnamefont {A.}~\bibnamefont {De\'{a}k}}, \bibinfo {author}
  {\bibfnamefont {G.~M.}\ \bibnamefont {Stocks}}, \ and\ \bibinfo {author}
  {\bibfnamefont {L.}~\bibnamefont {Szunyogh}},\ }\href {\doibase
  10.1088/0953-8984/26/18/186001} {\bibfield  {journal} {\bibinfo  {journal}
  {J. Phys.: Condens. Matter}\ }\textbf {\bibinfo {volume} {26}},\ \bibinfo
  {pages} {186001} (\bibinfo {year} {2014})}\BibitemShut {NoStop}%
\bibitem [{\citenamefont {Dup\'{e}}\ \emph {et~al.}(2014)\citenamefont
  {Dup\'{e}}, \citenamefont {Hoffmann}, \citenamefont {Paillard},\ and\
  \citenamefont {Heinze}}]{dupe_tailoring_2014}%
  \BibitemOpen
  \bibfield  {author} {\bibinfo {author} {\bibfnamefont {B.}~\bibnamefont
  {Dup\'{e}}}, \bibinfo {author} {\bibfnamefont {M.}~\bibnamefont {Hoffmann}},
  \bibinfo {author} {\bibfnamefont {C.}~\bibnamefont {Paillard}}, \ and\
  \bibinfo {author} {\bibfnamefont {S.}~\bibnamefont {Heinze}},\ }\href
  {http://www.nature.com/doifinder/10.1038/ncomms5030} {\bibfield  {journal}
  {\bibinfo  {journal} {Nat. Commun.}\ }\textbf {\bibinfo {volume} {5}},\
  \bibinfo {pages} {4030} (\bibinfo {year} {2014})}\BibitemShut {NoStop}%
\bibitem [{\citenamefont {R\'{o}zsa}\ \emph
  {et~al.}(2016{\natexlab{a}})\citenamefont {R\'{o}zsa}, \citenamefont
  {De\'{a}k}, \citenamefont {Simon}, \citenamefont {Yanes}, \citenamefont
  {Udvardi}, \citenamefont {Szunyogh},\ and\ \citenamefont
  {Nowak}}]{rozsa_skyrmions_2016}%
  \BibitemOpen
  \bibfield  {author} {\bibinfo {author} {\bibfnamefont {L.}~\bibnamefont
  {R\'{o}zsa}}, \bibinfo {author} {\bibfnamefont {A.}~\bibnamefont {De\'{a}k}},
  \bibinfo {author} {\bibfnamefont {E.}~\bibnamefont {Simon}}, \bibinfo
  {author} {\bibfnamefont {R.}~\bibnamefont {Yanes}}, \bibinfo {author}
  {\bibfnamefont {L.}~\bibnamefont {Udvardi}}, \bibinfo {author} {\bibfnamefont
  {L.}~\bibnamefont {Szunyogh}}, \ and\ \bibinfo {author} {\bibfnamefont
  {U.}~\bibnamefont {Nowak}},\ }\href {\doibase 10.1103/PhysRevLett.117.157205}
  {\bibfield  {journal} {\bibinfo  {journal} {Phys. Rev. Lett.}\ }\textbf
  {\bibinfo {volume} {117}},\ \bibinfo {pages} {157205} (\bibinfo {year}
  {2016}{\natexlab{a}})}\BibitemShut {NoStop}%
\bibitem [{\citenamefont {Hagemeister}\ \emph {et~al.}(2016)\citenamefont
  {Hagemeister}, \citenamefont {Vedmedenko},\ and\ \citenamefont
  {Wiesendanger}}]{Hagemeister}%
  \BibitemOpen
  \bibfield  {author} {\bibinfo {author} {\bibfnamefont {J.}~\bibnamefont
  {Hagemeister}}, \bibinfo {author} {\bibfnamefont {E.~Y.}\ \bibnamefont
  {Vedmedenko}}, \ and\ \bibinfo {author} {\bibfnamefont {R.}~\bibnamefont
  {Wiesendanger}},\ }\href {\doibase 10.1103/PhysRevB.94.104434} {\bibfield
  {journal} {\bibinfo  {journal} {Phys. Rev. B}\ }\textbf {\bibinfo {volume}
  {94}},\ \bibinfo {pages} {104434} (\bibinfo {year} {2016})}\BibitemShut
  {NoStop}%
\bibitem [{\citenamefont {Arblaster}(2010)}]{arblaster_crystallographic_2010}%
  \BibitemOpen
  \bibfield  {author} {\bibinfo {author} {\bibfnamefont {J.~W.}\ \bibnamefont
  {Arblaster}},\ }\href {\doibase 10.1595/147106710X493124} {\bibfield
  {journal} {\bibinfo  {journal} {Plat. Met. Rev.}\ }\textbf {\bibinfo {volume}
  {54}},\ \bibinfo {pages} {93} (\bibinfo {year} {2010})}\BibitemShut {NoStop}%
\bibitem [{\citenamefont {R\'{o}zsa}\ \emph {et~al.}(2015)\citenamefont
  {R\'{o}zsa}, \citenamefont {Udvardi}, \citenamefont {Szunyogh},\ and\
  \citenamefont {Szab\'{o}}}]{rozsa_magnetic_2015}%
  \BibitemOpen
  \bibfield  {author} {\bibinfo {author} {\bibfnamefont {L.}~\bibnamefont
  {R\'{o}zsa}}, \bibinfo {author} {\bibfnamefont {L.}~\bibnamefont {Udvardi}},
  \bibinfo {author} {\bibfnamefont {L.}~\bibnamefont {Szunyogh}}, \ and\
  \bibinfo {author} {\bibfnamefont {I.~A.}\ \bibnamefont {Szab\'{o}}},\ }\href
  {\doibase 10.1103/PhysRevB.91.144424} {\bibfield  {journal} {\bibinfo
  {journal} {Phys. Rev. B}\ }\textbf {\bibinfo {volume} {91}},\ \bibinfo
  {pages} {144424} (\bibinfo {year} {2015})}\BibitemShut {NoStop}%
\bibitem [{\citenamefont {R\'{o}zsa}\ \emph
  {et~al.}(2016{\natexlab{b}})\citenamefont {R\'{o}zsa}, \citenamefont {Simon},
  \citenamefont {Palot\'{a}s}, \citenamefont {Udvardi},\ and\ \citenamefont
  {Szunyogh}}]{rozsa_complex_2016}%
  \BibitemOpen
  \bibfield  {author} {\bibinfo {author} {\bibfnamefont {L.}~\bibnamefont
  {R\'{o}zsa}}, \bibinfo {author} {\bibfnamefont {E.}~\bibnamefont {Simon}},
  \bibinfo {author} {\bibfnamefont {K.}~\bibnamefont {Palot\'{a}s}}, \bibinfo
  {author} {\bibfnamefont {L.}~\bibnamefont {Udvardi}}, \ and\ \bibinfo
  {author} {\bibfnamefont {L.}~\bibnamefont {Szunyogh}},\ }\href
  {http://link.aps.org/doi/10.1103/PhysRevB.93.024417} {\bibfield  {journal}
  {\bibinfo  {journal} {Phys. Rev. B}\ }\textbf {\bibinfo {volume} {93}},\
  \bibinfo {pages} {024417} (\bibinfo {year} {2016}{\natexlab{b}})}\BibitemShut
  {NoStop}%
\end{thebibliography}
\end{document}